\documentclass[aip,graphicx,amsmath,amssymb,reprint,floatfix]{revtex4-1}

\usepackage[utf8]{inputenc}
\usepackage[T1]{fontenc}
\usepackage{graphicx}
\usepackage{dcolumn}
\usepackage{bm}
\usepackage{newtxtext, newtxmath}  
\DeclareGraphicsExtensions{.pdf,.jpeg,.jpg, .png, .eps, .tiff}
\usepackage{epstopdf}
\usepackage{lipsum}
\usepackage{tabularx}
\usepackage{booktabs}
\usepackage{array}
\usepackage{natbib}
\usepackage{xcolor}
\usepackage{grffile}
\usepackage{boxhandler}
\usepackage{float}
\usepackage[normalem]{ulem}
\usepackage{hyperref}

\draft 

\begin{document}


\title{Active particles knead three-dimensional gels into open crumbs} 



\author{Martin Cramer Pedersen}
\thanks{These authors contributed equally}
\affiliation{Niels Bohr Institute, University of Copenhagen, Denmark}

\author{Sourav Mukherjee}
\thanks{These authors contributed equally}
\affiliation{UGC-DAE Consortium for Scientific Research, University Campus, Khandwa Road, Indore 452017, India}

\author{Amin Doostmohammadi}
\affiliation{Niels Bohr Institute, University of Copenhagen, Denmark}

\author{Chandana Mondal}
\affiliation{UGC-DAE Consortium for Scientific Research, University Campus, Khandwa Road, Indore 452017, India}

\author{Kristian Thijssen}
\email[]{kristian.thijssen@nbi.ku.dk}
\affiliation{Niels Bohr Institute, University of Copenhagen, Denmark}


\date{\today}

\begin{abstract}
Colloidal gels are prime examples of functional materials exhibiting disordered, amorphous, yet meta-stable forms. They maintain stability through short-range attractive forces and their material properties are tunable by external forces.
Combining persistent homology analyses and simulations of three-dimensional colloidal gels doped with active particles, we reveal novel dynamically evolving structures of colloidal gels. Specifically, we show that the local injection of energy by active dopants can lead to highly porous, yet compact gel structures that can significantly affect the transport of active particles within the modified colloidal gel.
We further show the substantially distinct structural behaviour between active doping of 2D and 3D systems by revealing how passive interfaces play a topologically different role in interacting with active particles in two and three dimensions. 
The results open the door to an unexplored prospect of forming a wide variety of compact but highly heterogeneous and percolated porous media through active doping of 3D passive matter, with diverse implications in designing new functional materials to active ground remediation. 
\end{abstract}


\maketitle 



Many compounds in nature do not exhibit organised crystalline formations; instead, they appear as disordered, amorphous solids \cite{gupta1996non}. 
Colloidal gels are a quintessential example of this category of materials \cite{royall2021real},  maintaining a meta-stable state within local energy minima \cite{zaccarelli2007colloidal}.
This stability arises from slow dynamics emerging from short-range attractive forces between particles \cite{baxter1968percus, lu2008gelation}.
As a result, their mechanical properties depend on their history as the materials traverse the energy landscape due to ageing \cite{zaccarelli2007colloidal, tsurusawa2019direct, fielding2000aging, royall2021real}, or due to design protocols dictated by annealing \cite{pollard2022yielding} or externally applied forces \cite{zaccarelli2007colloidal, poon2002physics, cipelletti2005slow, masschaele2009direct, sprakel2011stress, grenard2014timescales, landrum2016delayed, johnson2018yield, koumakis2015tuning, nicolas2018deformation} which help overcome kinetic barriers \cite{patinet2016connecting}.

This energy injection does, however, not need to come from a global or external source. 
The advance of active particles \cite{wysocki2014cooperative, bechinger2016active}, e.g. self-propelled or swimming particles on the colloidal scale \cite{wang2013small, wang2015one, dey2017chemically}, has created the opportunity to regulate dynamics at the local scale due to internal forces/local energy injection \cite{palacci2013photoactivated, gao2014environmental, szakasits2019rheological}.
Much material design has emphasised the resulting steady states of large-scale active particle dynamics, which can result in dynamics not described by thermodynamics. 
These active environments affect the properties of passive particles, e.g. diffusion and viscosity are both modified in active baths \cite{hinz2014motility, takatori2015theory, wittkowski2017nonequilibrium, wang2019interactions, szakasits2019rheological}, and active systems can induce effective, attractive forces and repulsion in colloidal systems \cite{mallory2020universal, omar2018swimming}, possibly resulting in active clusters \cite{ginot2018aggregation} or active mixtures \cite{wysocki2016propagating, redner2013reentrant}.

\begin{figure}[!tb]
	\includegraphics[width=7.5cm]{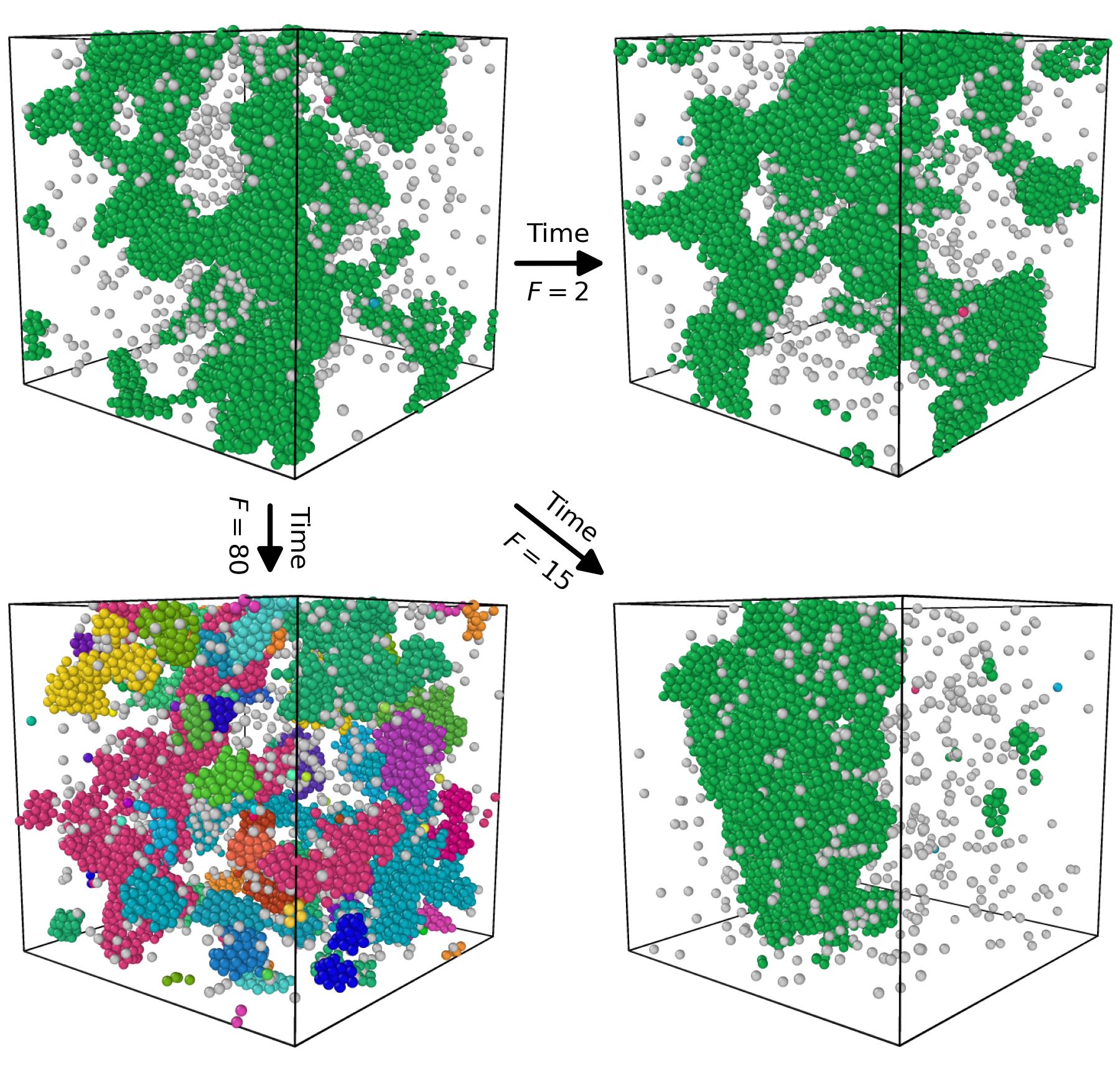}
	\caption{
 Varying the magnitude of the active force $F$ alters the manner in which the active particles knead the gel. White particles are active Brownian particles; the gel particles are color-coded according to the gel particle cluster to which they belong. From the starting configuration in the top left, we obtain the three depicted structures by running our simulations with the shown levels of activity.
 }
	\label{Figure:ABP_snaps}
\end{figure}

In diverse natural setups, active systems tend to inhabit complex amorphous surroundings; striking examples include self-propagating cells like swarming-motility of soil-dwelling \textit{M. xanthus}\cite{zusman2007chemosensory}, infiltration of \textit{E. coli} into leaf stomata\cite{ranjbaran2020mechanistic} or pathogenic \textit{S. typhimurium}/\textit{B. subtilis} in colonic/cervical mucus\cite{furter2019mucus}. Besides chemical interaction, these active particles can mechanically interact with their surroundings, bumping into surrounding media, deforming \cite{wang2019shape} and possibly penetrating it \cite{krakhmal2015cancer}. 
These kinetic interactions can affect both the surface and bulk reorganization of the amorphous media. 
Hence, the introduction of a few active particles in an amorphous passive medium, \textit{active doping}, \cite{omar2018swimming, wei2023reconfiguration}, can be used to reach thermodynamically favoured steady states of the complex passive surroundings. 
This can lead to the formation of crystalline structures \cite{ni2014crystallizing, das2019active}, regulate the structure of gels and glasses \cite{ni2013pushing} or transform the shape of vesicles \cite{wang2019shape}.

However, until now, most theoretical research on gels with active particles has focused on embedded particles, resulting dominantly in modifications of bulk properties \cite{szakasits2019rheological, mallory2020universal,omar2018swimming, massana2018active}. This is appropriate for 2D systems, where the topological loops in the gel network only serve as confined regions for the active particles. For a particle to propagate from one confined region to another, the active particles must penetrate the gel strands~\cite{borba2020controlling, massana2018active, mallory2020universal} or they will accumulate at solid surfaces \cite{van2008dynamics, elgeti2013wall, reichhardt2021clogging}, which can lead to interface sorting \cite{volpe2011microswimmers}.
In contrast, the topology of the gel network functions fundamentally differently in 3D. 
Enclosed regions are cavities, which may be absent in certain gels. The loops formed by the strands of the gel now act as archways, which are continuously modified, created and destroyed as active particles can quickly propagate through them without directly disrupting the connectivity of the gel. This is in contrast to stationary porous media with active particles, where the particles move around curved surfaces, hopping from one interface to the next \cite{bhattacharjee2019bacterial, moore2023active}.
These topological distinctions in the transient behaviour have not yet been addressed in evolving passive surroundings due to a lack of appropriate tools. 

Here, we demonstrate the significantly impact of active particles on the evolution of the complex surrounding medium in 3D, where interactions are dominated by the time scale of active particles and the relaxation time of the gel. 
Specifically, we reveal an optimal range of activity versus gel relaxation for which active dopants knead colloidal gels into open but compact three-dimensional networks.

\noindent{\bf Simulation methods.} We perform Langevin dynamics simulations using a model gel former~\cite{GTR2017}, consisting of $7000$ colloidal particles interacting with a Morse potential that mimics the short-range attractive depletion forces found in colloid-polymer mixtures. This potential is choses as it agrees well with the Asakura–Oosawa idealization of colloid–polymer mixtures \cite{taffs2010structural}. We consider a distribution of seven gel particles of varying size to suppress crystallization, with average diameter $\sigma$ and cut-off interaction distance $r^c_i$ for particle type $i$. All particles have unit mass density. Details of the model system and simulations can be found in the Supporting Materials (SM). Due to the short interaction range and strong attraction, this produces a stable, percolating gel on large time scales, as seen in Fig.~\ref{Figure:ABP_snaps}, with most reconfigurations occurring at the surface with surface diffusion time $\tau^p$ of $10^3$ Brownian times \cite{thijssen2023necking}. All simulations are done at volume fraction $\phi=0.08$. This volume fraction was chosen as it is near the coexistence curve of the dilute gas-gel boundary \cite{royall2015probing}. Active particles were introduced after the percolating gel had formed. We have tested that all our results are robust for different initialization conditions.

\noindent{\bf Doping with active Brownian particles.}
We consider the effect of active doping on a passive 3D gel, with the addition of $10\%$ active particles to the system. 
We model the active particles with only the repulsive part of the Morse potential, and active Brownian dynamics achieved by the introduction of a self-propulsion force ${\bf F_i} = F {\bf e_i}$, with magnitude $F$ and acting in the direction of the unit vector ${\bf e_i}$ that has a rotational diffuses with diffusion constant $D^r$.
The dynamics of a single ABP is mainly controlled by a dimensionless number, the \emph{P{\'e}clet} number $Pe = F\sigma / k_BT$~\cite{omar2018swimming}, defined as the ratio between advective and diffusive transport, with $D^t$ is the translational diffusion. Thereby providing a measure for the strength of the self-propulsion of the particles, scaling with active force $F$.


For low $F$, the gel evolution remains similar to the case without active particles present (Fig.~\ref{Figure:ABP_snaps} and Movie 1). For high activities, the gel breaks up into small clusters, which coarsen into sizes on length scales larger than the typical radius of a gel strand (Movie 2). 

Interestingly, for intermediate activities $F=15-20$, the full gel coarsens and is kneaded into a large clump (Fig.~\ref{Figure:ABP_snaps} and Movie 3). However, driven by these local forces, the passive system fails to attain its energy minima, characterized by a phase-separated solid object surrounded by the active gas. Instead, the gel retains large amounts of holes through which the active particles continuously pass (Movie 4).



\begin{figure}[!tb]
	\includegraphics[width=7.5cm]{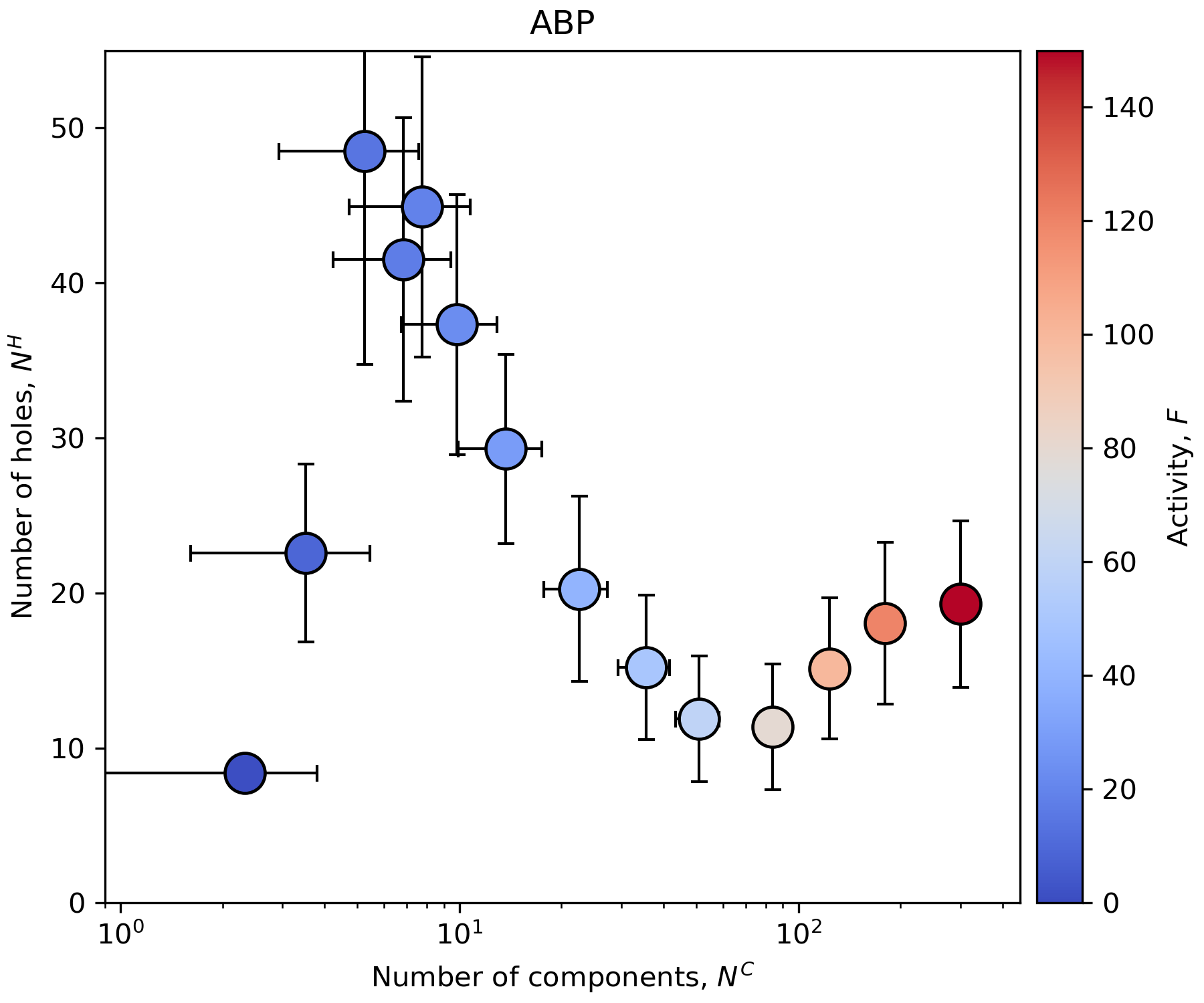}
	\caption{Tunable and non-monotonic topological structure of 3D colloidal gels with increasing activity of the dopants. The topological structure is characterized by the mean $\pm$ sd of number of connected components $N^C$ and the number of holes $N^H$ computed from the coordinates of the gel particles for simulation trajectories with Active Brownian Particles (ABP) upon reaching the steady state. We demonstrate and discuss the convergence of these quantities in the SI.}
	\label{Figure:Figure2}
\end{figure}
\begin{figure*}[!tb]
	\includegraphics[width=15.2cm]{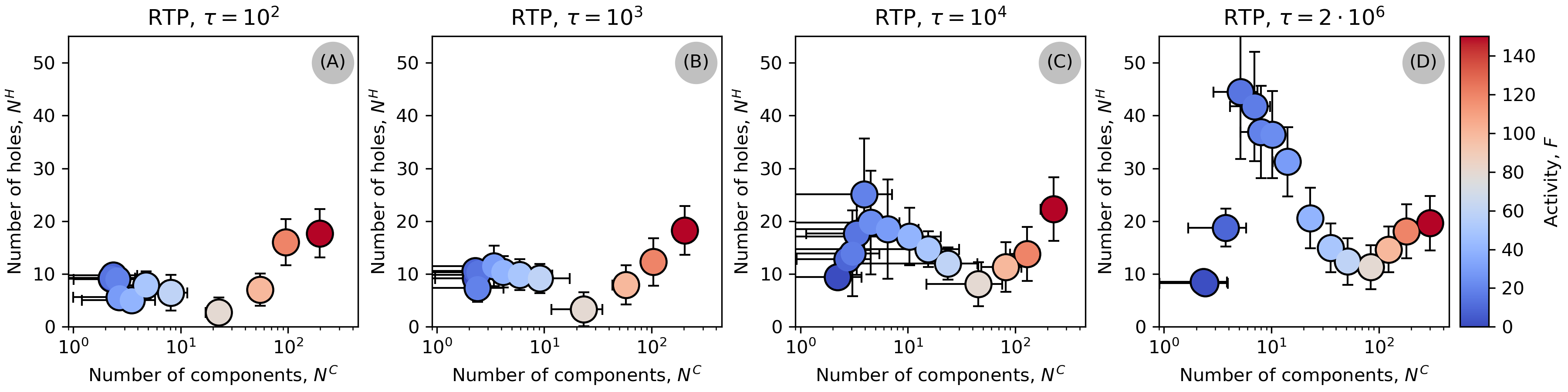}
	\caption{Tunable and non-monotonic topological structure of 3D colloidal gels with increasing activity $F$ of the Run-and-Tumble particles (RTP) for varying tumbling time scales $\tau$. The topological structure is characterized by the mean $\pm$ standard deviation of number of connected components $N^C$ and the number of holes $N^H$ computed from the coordinates of the gel particles for simulation trajectories.}
	\label{Figure:Figure3}
\end{figure*}
Unlike 2D, a detailed characterization of the 3D gel network presents several complexities; in particular in capturing the role and behaviour of the archway forming gel strands. To overcome this challenge, we use 
\emph{Topological data analysis} (TDA) 
to quantify these topological properties of the gel by probing the evolution of the mesoscale structure of our gels. The method allows us to track topological changes in our simulations over time using the mathematical language of \emph{persistent homology} (PH)~\cite{Verri1993, Robins1999, Edelsbrunner2002}. Using this language, we determine archetypal topological gel properties and how active particles impact these. PH is frequently used to analyze soft matter simulations and structures, e.g. amorphous materials, quasicrystals, protein compressibility, carbon allotropes, and polymer melts~\cite{Hiraoka2016, Pedersen2020, Gameiro2015, Xia2015, Shimizu2021}.
We compute the periodic alpha-shape filtration~\cite{Edelsbrunner1995, Edelsbrunner2010} of the points describing our gel in each simulation frame and analyze the homology groups of these complexes and their relationships with each other. The $k$th homology group for a given topological space encodes information on the topological $k$-features, the $0$-features being connected \emph{components} of the space and the $1$-features being the rings or \emph{loops} of the space making up the archways. We shall not discuss higher-order features like the enclosed \emph{cavities} ($k = 2$) here, as we focus on volume fractions for which these play little role. In the SI, we introduce the basics of PH and its data structures and cover in detail how we mathematically utilize this framework to define the number of holes, $N^H$, and the number of connected components, $N^C$, in the evolving morphology of the gel. We compute these two topological quantities for each frame in our trajectory - and in order to assess the general trend in the simulation, we display the mean of the new steady-state after the introduction of active particles (Fig. \ref{Figure:Figure2}). See SM for a discussion on gel time-evolution. 

Remarkably, using this method, we find that the number of archways initially increases rather than decreases upon the introduction of activity, which is what one could expect from equilibrium coarsening dynamics. This is in stark contrast to the 2D case, where activity just acts as an effective repulsive or attractive force, highlighting the difference in topological dimensionality. In 2D, the active particles are trapped in their respective confined cavities \cite{omar2018swimming}. Hence, in order to hop from one cavity to the next, they must push all the surface and bulk passive particles together, decreasing the number of holes.

In 3D, active particles not embedded in the gel move on passive interfaces, reorganizing them until active particles escape, which they can more easily do by forming large amounts of archways. Only for really high activities does the gel get bombarded with enough kinetic energy for strands to continuously break up: the gel forms small intermediate clusters, the size of which is dependent on the active particle density. This naturally destroys the holes/archways of the gel while increasing the number of components, which is quantitatively captured using persistent homology as shown in Fig.~\ref{Figure:Figure2}.

\noindent{\bf Doping with run-and-tumble particles.}
To underscore the interaction between active forces and passive surfaces, we incorporate tumbling mechanics into the behavior of active particles. This adjustment enables particles to escape passive interfaces more readily through random reorientations. Thus, instead of altering rotational and translational diffusion, we implement a run-and-tumble (RTP) motion, simulating the non-equilibrium directional shifts occurring within a time scale $\tau$. We implement the run-and-tumble motion of the ABP with in-house modifications to LAMMPS~\cite{PLIMPTON19951}. We make ${\bf e_i}$ time-dependent by changing its orientation randomly with a time interval on Poisson distributed intervals with mean $\tau$. 

This allows us to vary the effective persistent length of the active particles, as shorter tumbling times cause the particle to change orientation more often and, hence, allow active particles to reorganize away from the interface. For high tumbling times, we recover the ABP behaviour (Fig.~\ref{Figure:Figure3}(D) and movie 5). However, for lower tumbling times, we find that the duration of the contact between active particles and the surface is not sufficiently long to break the gel strands. For intermediate activity, the active particles are not in contact with the surface of the gel long enough to affect the diffusion of the surface gel particles. Hence, we find that the active particles cannot knead the gel into a compact form. This manifests in a lower number of holes $N^H$ as the $\tau$ decreases (Decrease of the peak in Fig.~\ref{Figure:Figure3}). For high activities, the active particles cannot break up the individual strands in this low $\tau$ limit. Rather than breaking the gel into small clusters, collisions merely knock off individual gel particles, resulting in a larger amount of individual gel particles in the gas phase, in coexistence with a large gel structures (movie 6) as seen from the structure factor analysis (see SM).

\begin{figure*}[!tb]
	\includegraphics[width=15.2cm]{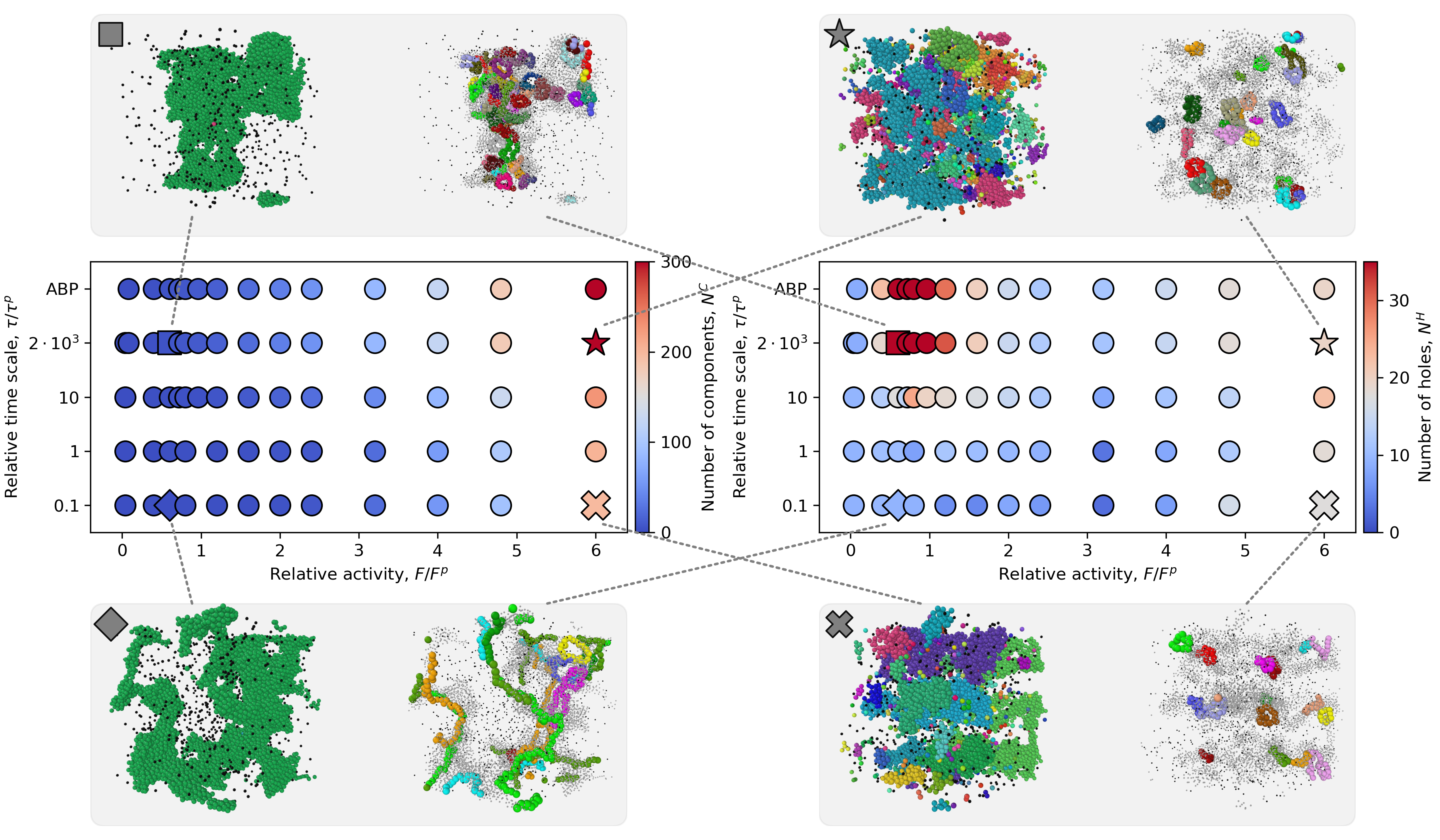}
	\caption{Phase diagrams of the topological structure of colloidal gels doped with active particles. In the middle row on the left, number of connected components $N^C$ is shown as a function of the activity type and magnitude; and on the right the number of holes $N^H$ under the same conditions. Above and below are snapshots of meta-stable frames in selected simulations highlighting the differences in topological structure, that gels exhibit throughout the phase diagram. We depict each frame twice; in the rendering on the left, each component is shown in a different color, whereas on the right, representatives of each hole are shown in different colors, specifically, \emph{stable volumes}~\cite{Obayashi2022}}
	\label{Figure:PhaseDiagram}
\end{figure*}
%


The crosstalk between activity and relaxation time and its impact on the gel structure is best represented in the phase diagrams in Fig.~\ref{Figure:PhaseDiagram}, where we normalize our axes by the magnitude of the attractive force between the gel particles and the relaxation time scale of the gel. Surprisingly, the diagrams show that the active particles do not purely set the dimensionless numbers that govern the gel evolution. 
Previous work~\cite{omar2018swimming} has shown that the active force of active particles embedded in the gel needs to be larger than the attractive forces $F^p$ holding the gel together. This leads to active particles escaping the gel, which causes the gel to reorganize as the active particles can cause effective attraction, resulting in phase-separated states, or repulsion, due to an increase in effective temperature.

Despite not having active particles embedded in the gel, we do find that any steady-state change caused by active particles' behaviour requires active forces larger than the attractive forces $F^p$ holding the gel together. We approximate the attractive forces as  $F^p\approx D^0/d^0\approx 25$ where $D^0$ is the depth of the Morse potential, and $d^0$ is the dissociation range, estimated where the Morse potential is $5\%$ above its minimum value. 
The transition from negligible impact to observable configurational changes in the gel as a function of activity is anticipated. Insufficient active forces will result in the gel particles maintaining their existing configuration, as cohesive forces will keep the structure intact.

Remarkably, the phase diagram reveals that the kneading behaviour only emerges when the surface relaxation time, estimated as $\tau^P\approx 10^3$ Brownian times for these gels  \cite{thijssen2023necking}, is smaller than the reorganization time of the active particles $\tau$  (Fig.~\ref{Figure:PhaseDiagram}).
When the active forces are within a moderate range, preventing immediate detachment of gel particles from the cluster due to their kinetic energy, we observe a phenomenon where active particles induce reorganization within the gel as they traverse passive interfaces. This reorganization leads the gel to descend to lower energy states within the landscape. However, owing to the dimensionality of the system, active particles consistently generate small apertures, facilitating transport through the passive object.

To capture this, we split the phase diagram into the number of components ($N^C$) and the number of holes ($N^H$) of the persistent homology measure. A traditional measurement of the number of connected components would be able to automatically detect the large activity region of the gel breaking up, which is only dependent on the active force compared to the attractive force $F/F^P$. Here, however, we show how the first component quantitatively captures the kneading behaviour, resulting in a system with a rich topology for high $\tau/\tau^P$. It is noteworthy that this behaviour is completely missed if one uses the standard method of analyzing only the number of clusters.

\noindent{\bf Discussion.}
Our results present one of the first studies of activity-induced reconfiguration of 3D colloidal gels. 
In the presence of active Brownian particles with polar driving, we find that the gel remains stable within the simulation time window when the self-propulsion speed (activity) is low. For high activities, the gel is driven apart and becomes unstable. 
However, for intermediate activities, the gel starts to be kneaded, being driven to a lower-energy state purely from surface interactions but never reaching its lowest-energy state, retaining large amounts of small archways to facilitate the surface transport of active particles. 
Consequently, we demonstrate that run-and-tumble particles with adjustable persistence lengths\cite{khatami2016active, saintillan2023dispersion} exhibit a behavior determined not only by the active force but also by the balance between the reorganization/tumbling time of active particles and the relaxation time of gel particles on the surface. 
By employing TDA, we further characterize the mechanism for formation of such an open network structure based on the interplay between the activity-induced time scale and the relaxation time of the colloidal gel.  

Our findings underscore the significance of distinguishing between 2D and 3D systems of gels and active particles. We have highlighted the stark differences in the topological descriptions of these systems, shedding light on the dynamics and structural behaviors unique to the active doping of 3D colloidal gels. In 2D, the interplay between active particles and gel interfaces is primarily governed by collision and penetration dynamics, profoundly affecting bulk diffusion\cite{volpe2011microswimmers}. On the contrary, in 3D, active particles predominantly diffuse around interfaces, influencing surface diffusion and leading to a more complex phase diagram of active doping. This contrast may elucidate the challenges in achieving stability of Motility-Induced Phase Separation (MIPS) in 3D systems~\cite{wysocki2014cooperative}. 

The application of TDA methods for quantifying gel structures offers a new venue for comparing, e.g. data from confocal microscopic or X-ray tomographic methods to simulations such as the ones investigated here. This in turn holds promise for diverse applications, from understanding the behavior of soil-dwelling bacteria in reconfigurable porous soils to elucidating the transport mechanisms of cancer cells through complex vascular networks and extracellular matrices.

\section*{acknowledgement}
C. M. acknowledges funding from the Interaction fellowship and SERB (Ramanujan fellowship File No. RJF/2021/00012). A. D. acknowledges funding from the Novo Nordisk Foundation (grant No. NNF18SA0035142 and NERD grant No. NNF21OC0068687), Villum Fonden (Grant no. 29476), and the European Union (ERC, PhysCoMeT, 101041418). K.T. acknowledges from the European Research Council under the European Union’s Horizon 2020 research and innovation programme (Grant Agreement No 101029079). 
 Views and opinions expressed are however those of the authors only and do not necessarily reflect those of the European Union or the European Research Council. Neither the European Union nor the granting authority can be held responsible for them.

\bibliography{bibliography}

\clearpage

\section{Supporting Materials}

\subsection{Movie captions}

Movie 1. The evolution of the gel after the introduction of active Brownian particles with active force $F=2$. Gel particles are coloured by cluster identification, similar to Fig. 1 in the manuscript. 

Movie 2. The evolution of the gel after the introduction of active Brownian particles with active force $F=100$. Gel particles are coloured by cluster identification, similar to Fig. 1 in the manuscript. 

Movie 3. The evolution of the gel after the introduction of active Brownian particles with active force $F=15$. Gel particles are coloured by cluster identification, similar to Fig. 1 in the manuscript. 

Movie 4. An individual hole was identified and tracked (orange circle) for the system with ABP (grey particles) and $F=15$. 

Movie 5. An individual hole was identified and tracked  (orange circle) for the system with RTP (grey particles) and $F=15$ and $\tau=2\cdot10^6$. 

Movie 6. The evolution of the gel after the introduction of Run and Tumble with active force $F=100$ and $\tau=100$. Gel particles are coloured by cluster identification, similar to Fig. 1 in the manuscript.

\subsection{Gel details and initialization}

We perform molecular dynamics simulations of a simple model gel~\cite{GTR2017} in which the particles $i$ and $j$ interact via truncated and shifted Morse potential $U(r_{ij})$: 
\begin{align}
	U(r_{ij}) = D^0 \left(e^{2\alpha(\sigma_{ij}-r_{ij})} - 2e^{\alpha(\sigma_{ij}-r_{ij})}\right),\,\ r<r^c_{ij}
	\label{Equation:Morse}
\end{align}

where, $r_{ij} = |{\bf r_j}- {\bf r_i}|$ with ${\bf r_i}$ being position of particle $i$, $\alpha=33$ is the range parameter, $D^0$ is the depth of the Morse potential, and $\sigma_{ij} = (\sigma_i + \sigma_j) / 2$, $\sigma_i$ being diameter of $i^{th}$ particle, with average size $\sigma$. We consider a poly-disperse additive mixture of particles of seven different sizes, with average size $\sigma$. The sizes are drawn from a Gaussian distribution of mean $\sigma$ and width $\delta$, with polydispersity $\delta/\sigma = 4\%$. The potential is truncated at the cutoff $r^c_{ij}$ and shifted to zero. This effectively reproduces the physics of colloid-polymer mixtures, leading eventually to gelation. 


The equation of motion for particle $i$ is given by the Langevin equation,
\begin{align}
	m_i\ddot{\bf r_i} = -\gamma \dot{\bf r_i} - {\bf \nabla_i}U + {\bf F_i}^{k_BT}
	\label{Equation:Langevin}
\end{align}
where $m_i$ is the mass and the “dots” denote derivatives with respect to time. Equations of motion are integrated with the Verlet algorithm with timestep $dt=0.002$ using LAMMPS. The friction coefficient $\gamma$ is chosen such that the dynamics is close to Langevin dynamics with Brownian time $\tau^B = (\sigma/2)^2/6D^t$ where $D^t$ is the translational self-diffusion constant for a particle. $D^t$ is related to $\gamma$ by Stokes's law $D^t=1/\beta\gamma$. ${\bf F_i}^{k_BT}$ denotes the delta-correlated thermal noise force acting on the $i^{th}$ particle with zero mean and variance $2k_BT\gamma/dt$, to fulfil the fluctuation–dissipation theorem. $T$ is the temperature and $U$ is interaction potential (Eqn.~\ref{Equation:Morse}). 


We express all quantities in dimensionless units, with length measured in units of $\sigma$, energies in units of the thermal energy $k_BT$, and time in units of $t = \gamma\sigma^2 / k T$.


We generate the initial configuration by placing $7000$ gel particles, with $1000$ particles of each type, placed randomly in a cubic box of lengths $Lx= Ly= Lz= L= 35.78$ at the desired volume fraction, which is defined as:
\begin{align}
    \phi = \frac{1}{6L^3} \sum_{i = 1}^{N} \pi \sigma_i^3
	\label{Equation:VolumeFraction}
\end{align}

All particles have unit mass-density. Initially random velocities are assigned to the particles from a Maxwell-Boltzmann distribution at inverse temperature $\beta$. Periodic boundary conditions are applied along all three directions. We choose $\gamma=20$, $\phi=0.08$, $D^0=5$, and $r^c_{ij}=1.4\sigma_{ij}$.

\subsection{Active particles}

\subsubsection{Active Brownian particles}

Once the percolating gel is generated, we introduce active Brownian particles (ABP) to the system with the aim of seeing whether or not the presence of active particles, which inject energy into the system, could make the gel age faster. The active particles have diameter $1$ and interact with each other and the gel particles with the Morse potential as given in Eqn. ~\ref{Equation:Morse}. In addition they are subjected to the active, self-propulsion force ${\bf F_i}$,
\begin{align}
        {\bf F_i} = F {\bf e_i}
        \label{Equation:ABP}
\end{align}
where $i$ is the particle the force is being applied to, $F$ is the magnitude of the force, and ${\bf e_i}$ is the vector direction of the force. We specify ${\bf e_i}$ via the components $(s_x,s_y,s_z)$ which are defined within the coordinate frame of the particle’s ellipsoid. We chose $s_x=1, s_y=0, s_z=0$ which sets the self-propulsion force to point along $x$-direction of the particle's body frame of axis. 

\subsubsection{Run and tumble}
To mimic the run and tumble motion of actual bacteria and study the effect of active directional change into the gel pattern, we implement run and tumble motion of the ABP with in-house modifications to LAMMPS. We call these particles with active direction change as Run and Tumble particles (RTP). The active propulsion force, in this case, has the form,
\begin{align}
        {\bf F_i} = F {\bf e_i(t)}
        \label{Equation:RT}
\end{align}
Note that, the direction of active force become time dependent such that the active force can change its direction on Poisson distributed intervals with mean $\tau$, so that $\tau$ controls the reversal frequency. The components $(s_x,s_y,s_z)$ of ${\bf e_i}$ are initialized with the value $s_x=1, s_y=0, s_z=0$ so that the active force initially points towards $x$-direction of the particle's body frame of axis similar to ABP. However, as time goes on, it can switch the direction on Poisson distributed intervals,  $\tau^p$, and $s_x,s_y,s_z$ can take any arbitrary value on the orientation sphere such that the self-propulsion force can point in any arbitrary direction of the particle's body frame of axis. The rotational dynamics of RTP can be controlled by an additional \emph{P{\'e}clet} number:
$Pe^{R}=\tau D^r$, with $\tau$ being the average time between tumble events and $D^r=3D^t/\sigma^2$ the rotational diffusion constant. $Pe^{R}$ determines if rotational degrees of freedom are dominated by tumbling events (small values) or by rotational diffusion (large values).

\subsection{Structure factors}

\begin{figure}[!htb]
	\includegraphics[width=8.5cm]{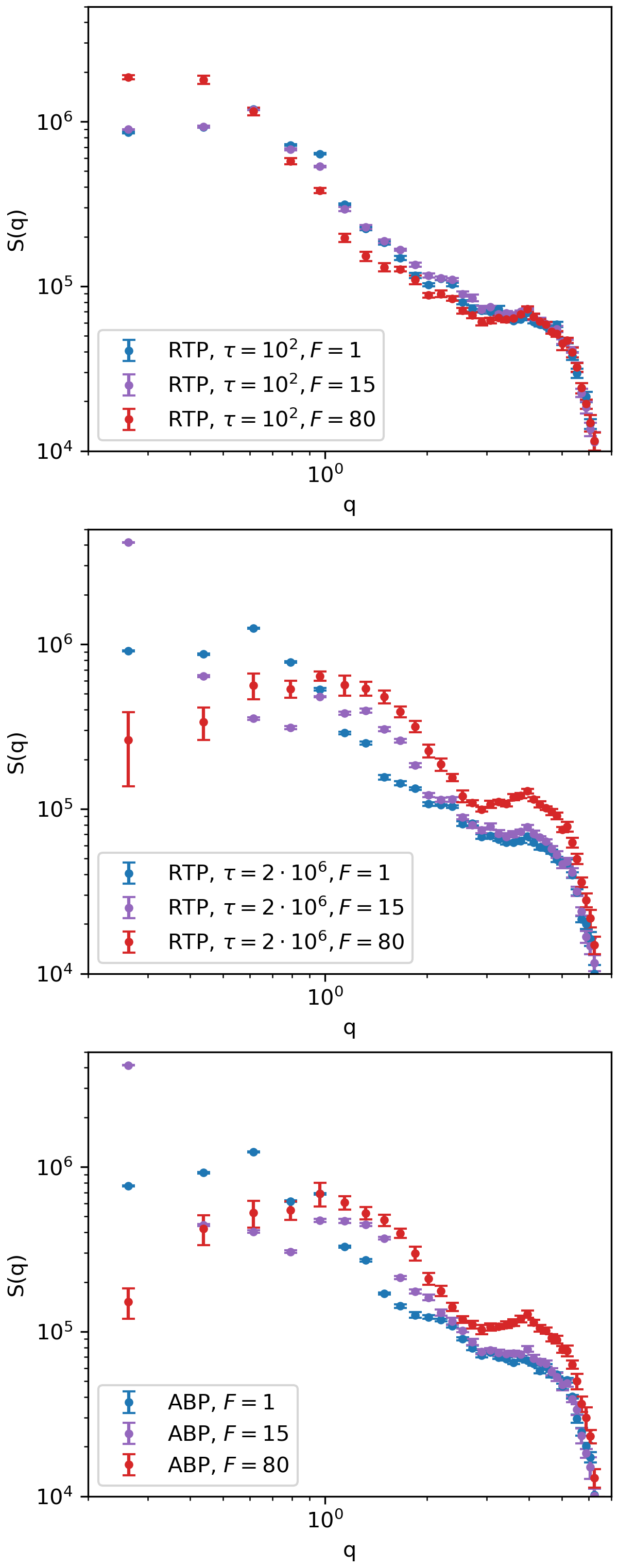}
	\caption{Averaged structure factor, $S(q)$, of the final $10$ frames in selected simulation trajectories.}
	\label{Figure:FigureSQ}
\end{figure}

To quantify the size of the structures emerging in the gel, we perform structural classifications on gels focusing on two-point correlations (static structure factor). One can identify emerging characteristic length scales from the structure factor $S(q)$, computed over $N$ particles directly in reciprocal space as:
\begin{align}
        S(q) = N^{-1} \sum_{i=1}^{N}\sum_{j\geq i}^{N} \langle e^{-i{\bf q \cdot r_{ij}}}\rangle
        \label{Equation:Sq}
\end{align}
where the average is performed on the ensemble of the $10$ final frames in our simulation trajectory (our frames are stored every $2\cdot10^5$ simulation steps) and evaluated isotropically at $q = |{\bf q}|$. The calculated structure factors are shown in Fig.~\ref{Figure:FigureSQ}.

We can observe three peaks that appear/disapear for varying activity and tumbling times. For low activities, we see that low q is prominent, indicating a percolating cluster, with a secondary peak at high $q$, which are individual particles.

For low tumbling times, we observe that low $q$ becomes more prominent with increasing $F$, indicating that the percolating gel network remains present, while the peak for high $q$ becomes a bit sharper as individual gel particles do get separated from the gel network.

For high $\tau$ and the ABP particles on the other hand, we observe that the low $q$ decreases as the percolating gel is broken up. Rather we now get a peak at medium $q$, corresponding to finite size clusters. 

\subsection{Persistent homology}

This section covers our topological approach to analysing the time series of our gels. As described in the main text, we deploy persistent homology to analyse each frame in our trajectory and quantify the topology of the gel in each frame.

\begin{figure*}[!tb]
	\includegraphics[width=17cm]{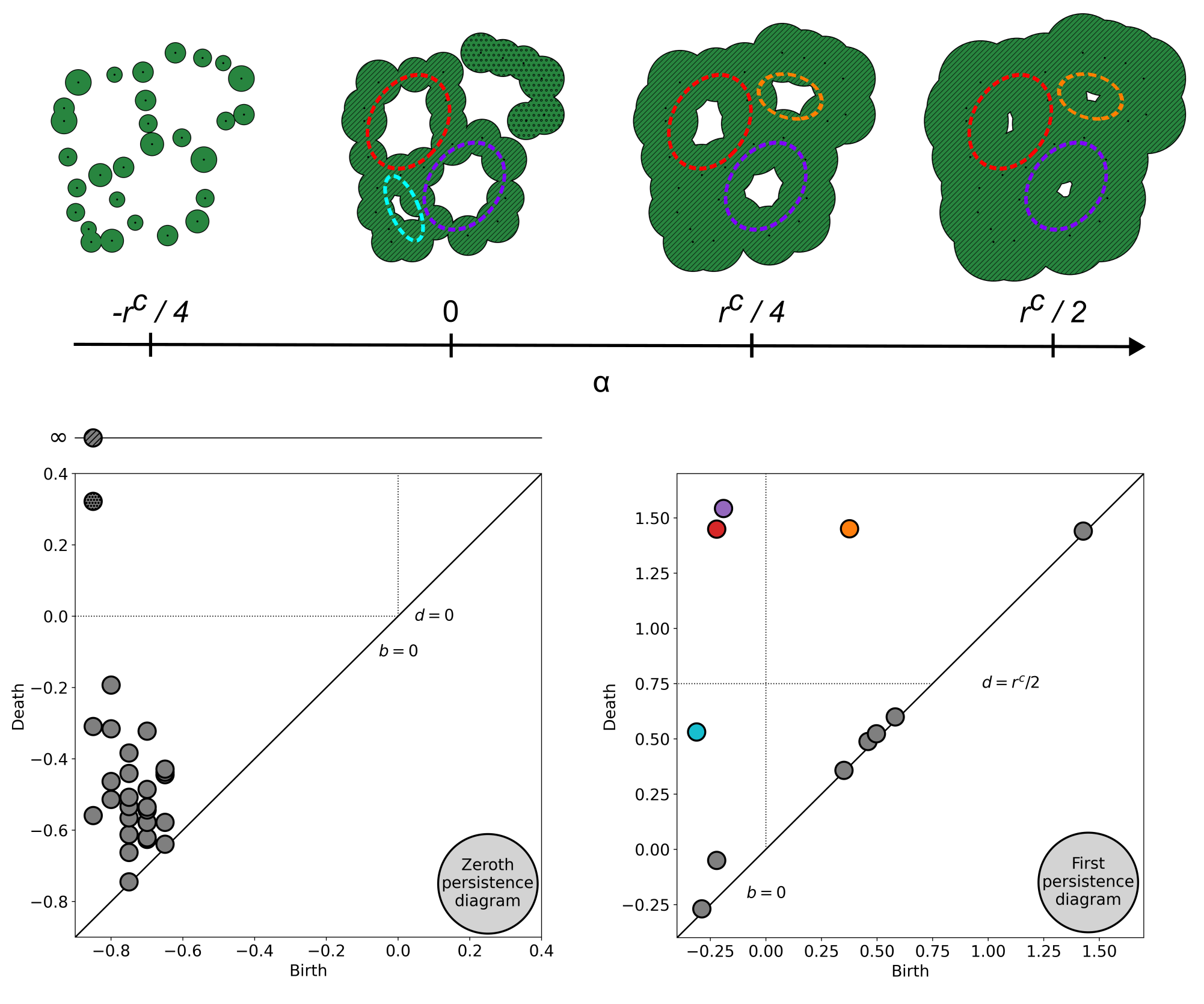}
	\caption{Alpha shape filtration of a weighted point set, shown on top, with the associated persistence diagrams, $PD0$ and $PD1$, shown below. For each value of $\alpha$, the point centers are imbued with a radius of their individual weight ($r^c / 2$ in our simulations) plus $\alpha$ producing another complex in the sequence. Selected topological features are highlighted and discussed in the text.}
	\label{Figure:AlphaShapes}
\end{figure*}

We construct \emph{persistence diagrams} ($PD$) encoding changes in homology of the alpha shape complexes in our alpha shape filtrations for each frame. The $k$th persistence diagram, $PDk$, is a collection of points describing how topological $k$-features emerge and disappear in our alpha shape filtration. In the context of PH, one makes use of the notion of the ``birth'', $b$, and the ``death'', $d$, of a topological feature to specify the lowest radius, for which a topological features appears, and the largest radius, after which it has disappeared. A point in the $k$th persistence diagram is simply an encoding of the birth and death $(b, d)$ of a $k$-feature in the given frame or point set. Additionally, one defines the persistence of a given features as the range of radii, for which it exists: $d - b$. The notion is sketched in Fig.~\ref{Figure:AlphaShapes}, where we see the first persistence diagram, $PD1$, associated to the point cloud also depicted in the figure.

As we are particularly interested in the percolation, porosity, and strand formation and breakage, we focus our attention on the first persistence diagrams, which encodes the loop structure of a given time frame in our simulation trajectory. These diagrams allow us to specify and quantify the notion of ``a hole''. We define the number of holes in a time frame of our simulations as a topological $1$-feature, for which:
\begin{align*}
	b \leq 0 \quad\quad d \geq r^c / 2
\end{align*}
where $r^c$ is the mean cutoff radius in the potentials of our particle ensemble. Reiterating, for a topological $1$-feature to be counted as a hole, we require that all constituent particles are within interaction distance of the neighboring points constituting the ring (the first criteria above), and that the hole is sufficiently large for another particle to pass through without interacting with any of the constituent particles (the second criteria above). Glancing at Fig.~\ref{Figure:AlphaShapes}, we observe that the topological features indicated by the red and purple loop satisfy these criteria, whereas the features indicated by the blue and orange do not. Specifically, the blue loop does not allow another particle to pass through it, i.e. its death is smaller than $r^c / 2$, and the orange loop is formed too late in the sequence, i.e. its birth is larger than $0$. This allows one to express the number of holes, $N_H$, in a given frame of our simulation as:
\begin{align}
	N_H = \sum_{(b_j, d_j) \in PD1} H(-b_j) H(d_j - r^c / 2)
	\label{Equation:NumberOfHoles}
\end{align}
where $H$ is the Heaviside function.

Additionally, we can introduce the notion of the number of components, $N_C$, in our gel simulations. As connectedness is encoded in the zeroth persistence diagram, $PD0$, we can readily introduce the notion of components in our gel in a fashion analogous to the number of holes; however, in this case we are simply counting topological features formed by particles interacting with each other. Expressed mathematically, we compute the following sum:
\begin{align}
	N_C = \sum_{(b_j, d_j) \in PD0} H(-b_j) H(d_j)
	\label{Equation:NumberOfComponents}
\end{align}
Using this construction, we can compute the number of components in the point set in Fig.~\ref{Figure:AlphaShapes} to be $2$; one with $(b, d)$ at approximately $(-0.85, 0.35)$ and one at approximately ($-0.85, \infty$). These are highlighted in Fig.~\ref{Figure:AlphaShapes} by different textures for $\alpha \geq 0$. Note that the latter is an example of a so-called \emph{essential feature}, meaning that it does not die for any value of $\alpha$; representing that our alpha shape complex is a single connected component values of $\alpha$ above $0.35$.

Thanks to recent developments, we can visualize the specific point constellations generating respective cycles by computing their \emph{stable volumes}~\cite{Obayashi2023}. See Fig.~\ref{Figure:PhaseDiagram} for an examples of these. Our persistent homology calculations were done using the software modules HomCloud~\cite{Obayashi2022} and Dionysus~\cite{Morozov2023}. We handle our molecular simulation trajectories using the Python module MDAnalysis~\cite{Michaud-Agrawal2011}.

\subsection{Convergence of topological quantities}

\begin{figure}[!htb]
	\includegraphics[width=8.5cm]{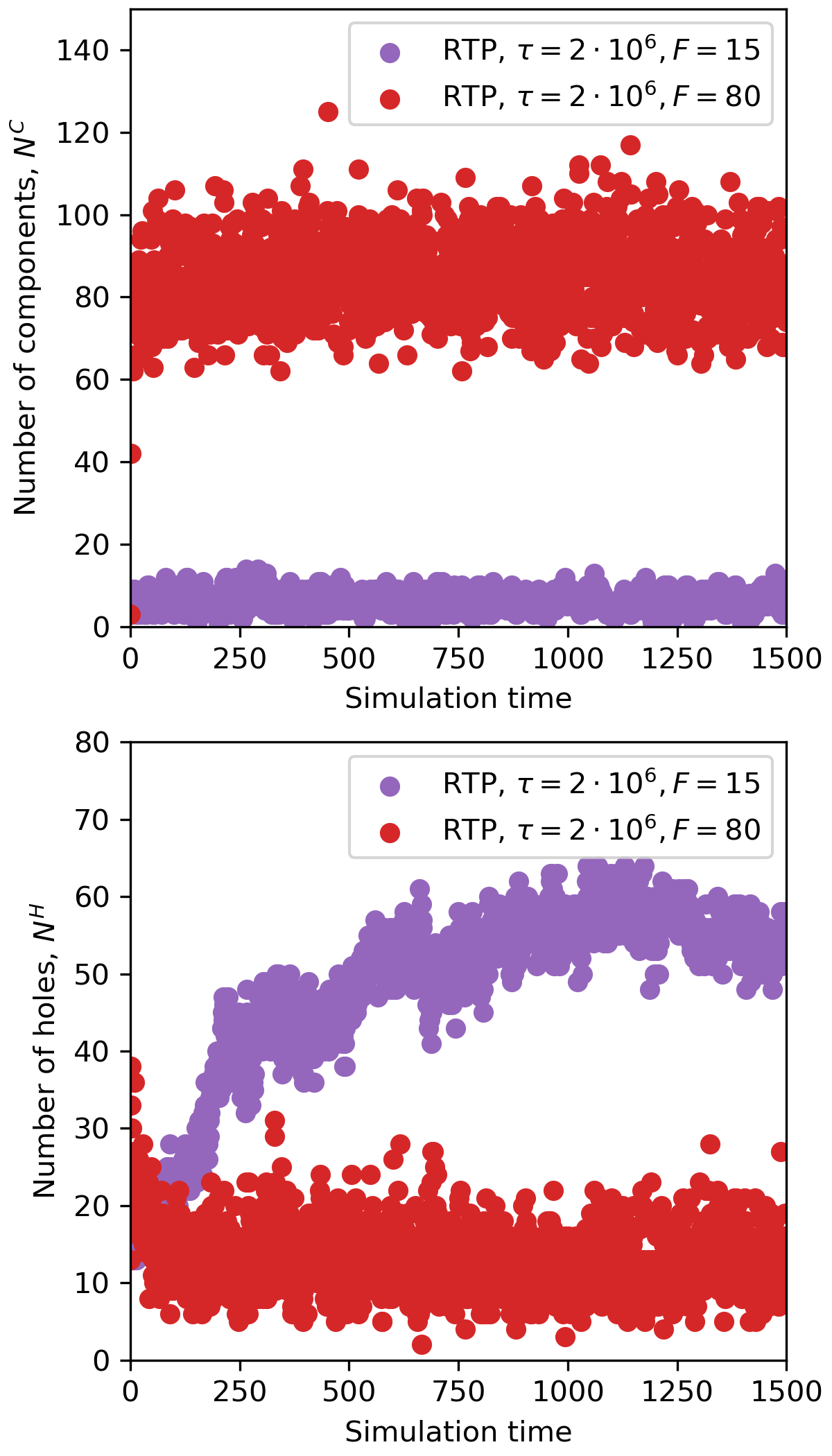}
	\caption{Time series of our topological quantities for selected simulations.}
	\label{Figure:Convergence}
\end{figure}

Figure~\ref{Figure:Convergence} illustrates the convergence of the topological quantities introduced in Eqns.~\eqref{Equation:NumberOfHoles} and~\eqref{Equation:NumberOfComponents}. We note that $N^C$ in particular appears to converge during the initial stages of our simulations.

\clearpage


%
%

%



\end{document}